\documentclass[prd,showpacs,showkeys,floatfix,nofootinbib,eqsecnum,fleqn,
               preprint,12pt,tightenlines]{revtex4} 

\usepackage{amsmath,amssymb,revsymb,graphicx,dcolumn}
\newcommand  {\version}{v6}

\newcommand{\beq}{\begin{equation}}
\newcommand{\eeq}{\end{equation}}
\newcommand{\beqa}{\begin{eqnarray}}
\newcommand{\eeqa}{\end{eqnarray}}
\newcommand{\bsubeqs}{\begin{subequations}}
\newcommand{\esubeqs}{\end{subequations}}

\newcommand{\dd}{\mathrm{d}}                    

\newcommand{\Eew}{E_\text{ew}}
\newcommand{\tew}{t_\text{ew}}
\newcommand{\EPlanck}{E_\text{Planck}}
\newcommand{\tPlanck}{t_\text{Planck}}

\begin{document}

\noindent Phys. Rev. D 80, 083001 (2009)   \hfill
          arXiv: 0905.1919 (\version)\newline\vspace*{2mm}

\title{Vacuum energy density kicked by the electroweak crossover\vspace*{5mm}}

\author{F.R. Klinkhamer}
\email{frans.klinkhamer@physik.uni-karlsruhe.de}
\affiliation{\mbox{Institute for Theoretical Physics, University of Karlsruhe (TH),}\\
76128 Karlsruhe, Germany}

\author{G.E. Volovik}
\email{volovik@boojum.hut.fi}
\affiliation{\mbox{Low Temperature Laboratory, Helsinki University of Technology,}\\
\mbox{Post Office Box 5100, FIN-02015 HUT, Finland}\\
and\\
\mbox{L.D. Landau Institute for Theoretical Physics, Russian Academy of Sciences,}\\
Kosygina 2, 119334 Moscow, Russia\\}

\begin{abstract}
\vspace*{2.5mm}\noindent
Using $q$--theory, we show that the electroweak crossover
can generate a remnant vacuum energy density
$\Lambda \sim \Eew^8/\EPlanck^4$,
with effective electroweak energy scale
$\Eew \sim 10^{3}\;\text{GeV}$
and reduced Planck-energy scale $\EPlanck \sim 10^{18}\;\text{GeV}$.
The obtained expression for the effective cosmological constant $\Lambda$
may be a crucial input for the suggested solution by Arkani-Hamed \emph{et al.}
of the triple cosmic coincidence puzzle
(why the orders of magnitude of the energy densities of vacuum, matter,
and radiation are approximately the same in the present Universe).
\end{abstract}

\maketitle

\section{Introduction}\label{sec:Introduction}

The $q$--theory description of the quantum vacuum provides
a natural cancellation mechanism for the vacuum energy
density~\cite{KlinkhamerVolovik2008a,KlinkhamerVolovik2008b,KlinkhamerVolovik2008c}.
The basic idea is to consider the macroscopic equations of a  \emph{conserved}
microscopic variable $q$, whose precise nature need not be known.
For a particular realization of $q$, it was found~\cite{KlinkhamerVolovik2008b}
that, if the vacuum energy density has initially a large Planck-scale value,
$\rho_{V}\sim \EPlanck^4$, it relaxes according
to the following power-law modulation:
\bsubeqs\label{eq:VacuumEnergyDecay}
\beqa
 \rho_{V}(t)\;\Big|^\text{nondissipative}
 \propto
 \frac{\omega^2}{t^2}\;\sin^2 \omega\,  t\,,
\label{eq:VacuumEnergyOscillating-dimensionfull}
\eeqa
with $\hbar=c=k=1$ in natural units and
a frequency $\omega$ of the order of the reduced Planck-energy scale
$\EPlanck \equiv 1/\sqrt{8\pi G_{N}} \approx
2.44\times 10^{18}\:\text{GeV}$. Quantum
dissipative effects have not been taken into account in the above result.
Indeed, matter field radiation (matter quanta emission)
by the oscillations of the vacuum can be expected to lead to
faster relaxation~\cite{Starobinsky1980,Vilenkin1985},
\beqa
 \rho_{V}(t)\;\Big|^\text{dissipative}
 \propto \Gamma^4  \exp(-\Gamma\, t)\,,
\label{eq:VacuumEnergyQuantum-dimensionful}
\eeqa
\esubeqs
with a decay rate $\Gamma\sim \omega \sim \EPlanck$.

In the present article, we consider what happens during the electroweak
crossover~\cite{Csikor-etal1999} of a spatially flat
Friedmann--Robertson--Walker (FRW) universe~\cite{Weinberg2008} at cosmic time
\beq
\tew\sim
\EPlanck/\Eew^2\,,
\label{eq:t_ew}
\eeq
where $\Eew\sim 10^{3}\:\text{GeV}$ is the effective electroweak
energy scale. In the epoch before the crossover, the vacuum energy
density has already relaxed to zero, according to
\eqref{eq:VacuumEnergyQuantum-dimensionful}. The classical equations of
$q$--theory demonstrate that during the epoch when only ultrarelativistic
matter (``radiation'') is present,
i.e., when the matter equation-of-state (EOS) parameter $w_{M}\equiv
P_{M}/\rho_{M}$ is exactly 1/3,  the vacuum energy density remains
strictly zero. But $w_{M}(t)$ deviates from 1/3 during the electroweak
crossover and the subsequent period when massive particles annihilate.
This implies, as will be shown in the present article, that
the vacuum energy density moves away from zero and acquires,
at $t \sim \tew$, a positive value of order
\beq \label{eq:vac_energy_at_ew}
\rho_{V,0}(t)  \sim \big( w_{M}(t) - 1/3 \big)^2\;H^4(t)\,,
\eeq
where the suffix 0 will be explained later and $H(t)$ is the Hubble parameter
of the spatially flat FRW universe considered.

After the electroweak crossover, the value $w_{M}=1/3$ is restored
and, if no other effects are operative,
the vacuum energy density smoothly returns to a zero value.
If, however, quantum relaxation effects are taken into account, the vacuum
energy density does not return to zero, but approaches a constant
value, which is of the order of the vacuum energy density
\eqref{eq:vac_energy_at_ew} at $t\sim \tew$.
This remnant vacuum energy density corresponds to the measured value of
the cosmological constant (see, e.g., Refs.~\cite{Weinberg2008,Komatsu2008}
and other references therein):
\beqa
\label{eq:vac_energy_at_present}
\Lambda &\equiv& \lim_{t \to\infty}\,\rho_{V}(t)
\sim  \rho_{V,0}(\tew)
\sim H^4(\tew)
          \sim 
\tew^{-4} \sim  \big( \Eew^2/\EPlanck\big)^4
\sim \big( 10^{-3}\:\text{eV} \big)^4  \,,
\eeqa
for the energy scales $\EPlanck$ and $\Eew$ defined
under \eqref{eq:VacuumEnergyOscillating-dimensionfull} and \eqref{eq:t_ew},
respectively.
The several steps in \eqref{eq:vac_energy_at_present} will be detailed
in the following, with the most important intermediate steps
collected in \eqref{eq:VacuumEnergySmoothEWtime} and \eqref{eq:FinalrhoV}.

The scenario outlined above differs from that of a cosmological phase
transition, for which the vacuum energy density may only decrease (changing
to a negative value if it was originally zero), and resembles the scenario in
which the vacuum energy density is generated by the conformal anomaly. In fact,
it has been suggested in Refs.~\cite{Schutzhold2002,KlinkhamerVolovik2009}
that the conformal anomaly of quantum chromodynamics (QCD)
gives rise to the vacuum energy density
$\rho_{V}(t)\propto |H(t)|\, E_\mathrm{QCD}^3$, where
$E_\mathrm{QCD} \sim 10^2\;\text{MeV}$
is the QCD energy scale (see also Ref.~\cite{Bjorken2004} for related remarks).
The rigorous microscopic derivation of this nonanalytic term has not yet been
given, as it requires the detailed behavior of QCD in the infrared.
For the moment, the main motivation of this particular nonanalytic
term is that it naturally provides the correct order of magnitude for the
present vacuum energy density and appears to give a good description of the
late evolution of the Universe~\cite{Klinkhamer2009}.
We remark also that part of the contribution
of the conformal anomaly to the vacuum energy density
has been estimated~\cite{Thomas-etal2009} as $\rho_{V}(t)\propto H^4(t)$,
which has the same $H$ dependence as \eqref{eq:vac_energy_at_ew}.
But the mechanisms of Ref.~\cite{Thomas-etal2009} and the present article
are different, as will be explained later.

The scenario with the emergence of a
positive vacuum energy density  \eqref{eq:vac_energy_at_present}
triggered by the electroweak crossover confirms the earlier suggestion
by Arkani-Hamed \emph{et al.}~\cite{ArkaniHamed-etal2000}
that electroweak physics is at the origin of a
``triple cosmic coincidence'' for the matter, radiation, and
vacuum energy densities in the present Universe
(see also the general discussion in Ref.~\cite{Chernin2008}).
While the coincidence among the matter and radiation energy densities appears to
be justified by the electroweak scenario~\cite{ArkaniHamed-etal2000},
the coincidence of these two ingredients
with the remnant vacuum energy density (effective cosmological constant) $\Lambda$
requires a particular relation in terms of the electroweak energy scale
$\Eew$  and the ultraviolet energy scale
$\EPlanck$, namely, $\Lambda  \sim    \Eew^8/\EPlanck^4$.
In order to explain this particular relation,
the authors of Ref.~\cite{ArkaniHamed-etal2000} suggested a
phenomenological model but had to assume
(page 4436, right column of the cited reference)
that ``an unknown mechanism canceled the vacuum energy density
at the global minimum of the potential.''
In our scenario, this mechanism is natural.

\section{Dynamical equations}
\label{sec:Dynamical_equations}

The present discussion starts from the theory outlined in
Ref.~\cite{KlinkhamerVolovik2008b}. We introduce a special conserved
quantity, the  vacuum ``charge''  $q$, to describe the statics and dynamics
of the quantum vacuum.
An example of this vacuum variable is given by the four-form field
strength~\cite{DuffNieuwenhuizen1980,Aurilia-etal1980,Hawking1984,Duff1989,
DuncanJensen1989,BoussoPolchinski2000,Aurilia-etal2004,Wu2008},
expressed in terms of $q$ as
$F_{\alpha\beta\gamma\delta}(x)$ $=$
$q(x)\,\sqrt{-g(x)}\, \epsilon_{\alpha\beta\gamma\delta}$.
But the dynamic equations for the vacuum variable $q$ and
the metric $g_{\alpha\beta}$
are \emph{universal}, that is, they do not depend on the particular realization of $q$. For
example, in the four-form realization, the generalized Maxwell equation for
the $F$--field is reduced to the following generic equation for the charge $q$:
\begin{equation}
\frac{\partial\epsilon(q)}{\partial q} + R\,\frac{\partial K(q)}{\partial q}=\mu\,,
\label{eq:genMaxwellSolution}
\end{equation}
where $\epsilon(q)$ is the vacuum energy density expressed in terms of $q$
[the possible dependence on other fields is kept implicit],
$R$ the Ricci curvature scalar,
$K(q)$ the gravitational coupling parameter which depends on the vacuum state,
and $\mu$  an integration constant.
The latter quantity $\mu$ plays the role of a Lagrange multiplier
related to the conservation of the charge $q$ and corresponds to the chemical
potential in thermodynamics~\cite{KlinkhamerVolovik2008a,KlinkhamerVolovik2008b}.

The metric field $g_{\alpha\beta}$  obeys the generalized Einstein equation
\bsubeqs\label{eq:genEinsteinEquation-rhoV}
\begin{eqnarray}
2K\,\big( R_{\alpha\beta}-g_{\alpha\beta}\,R/2 \big)&=&
-2\,\big(  \nabla_\alpha\nabla_\beta - g_{\alpha\beta}\, \square\big)\, K(q)
+\rho_{V}(q)\, g_{\alpha\beta} - T_{\alpha\beta} \,,
\label{eq:genEinsteinEquation}\\
\rho_{V}(q) &\equiv&  \epsilon(q)-  \mu\, q \,,
\label{eq:rhoV}
\end{eqnarray}
\esubeqs
where the metric has signature $(-,+,+,+)$ and $T_{\alpha\beta}$
is the matter energy-momentum tensor with vanishing covariant divergence
$\nabla_\alpha\,T^{\alpha\beta}=0$  from general coordinate invariance.
The particular combination \eqref{eq:rhoV}, and not $\epsilon(q)$,
is seen to determine the cosmological term in \eqref{eq:genEinsteinEquation},
which is perhaps the most important characteristic of our approach.

In what follows, we choose a value $\mu_0$ of the integration
constant $\mu$ in such a way that, in the absence of matter or other types
of perturbations, the solution of the equations corresponds to
the full-equilibrium Minkowski-spacetime vacuum.
The actual value $\mu_0$ and corresponding charge  $q_0$ of
the equilibrium vacuum  are determined by two equations:
\bsubeqs\label{eq:equil-eqs}
\beqa
\Big[\dd \epsilon(q)/\dd q-  \mu\,    \Big]_{\mu=\mu_0\,,\,q=q_0} &=&0\,,\\[2mm]
\Big[\epsilon(q)    -  \mu\, q\,\Big]_{\mu=\mu_0\,,\,q=q_0} &=&0\,,
\label{eq:equil-eqs-GDcondition}
\eeqa
\esubeqs
which follow from \eqref{eq:genMaxwellSolution}
and \eqref{eq:genEinsteinEquation-rhoV},
respectively, for $R_{\alpha\beta}=T_{\alpha\beta}=0$
and spacetime-independent $q_0$.
The equilibrium conditions \eqref{eq:equil-eqs} are supplemented by the
following stability condition:
\begin{equation}
\big(\chi_0\big)^{-1} \equiv
q^2\:\frac{d^2\epsilon(q)}{dq^2}\,\Bigg|_{q=q_0} > 0\,,
\label{eq:chi_0}
\end{equation}
where $\chi$ corresponds to the
vacuum compressibility~\cite{KlinkhamerVolovik2008a}.

The ``cosmological constant problem'' would be completely
solved if we could explain the origin
of this particular value $\mu_0$ for the integration constant $\mu$ appearing
in \eqref{eq:genMaxwellSolution} and \eqref{eq:genEinsteinEquation-rhoV}.
Here, our assumption is that the
Minkowski-spacetime vacuum is a \emph{self-sustained} system,
i.e., an isolated system that can exist without external pressure,
at $P=0$.  In general, the vacuum pressure $P$ and the vacuum
energy density $\epsilon$ are related by the thermodynamic Gibbs--Duhem
equation~\cite{KlinkhamerVolovik2008a}, $P=-\epsilon + \mu\, q$.
The vanishing pressure $P$ allowed for a self-sustained system
(from the assumed absence of external pressure)
then gives the additional condition \eqref{eq:equil-eqs-GDcondition},
which fixes $\mu$ to the value $\mu_0$.
From this viewpoint, cosmology corresponds to
the dynamic process of approach to the equilibrium state with $q=q_0$,
which is natural for any system isolated from the external environment.

Close to equilibrium, at $|q-q_0|\ll |q_0|$,
the dynamics of the system is determined by the coefficients
in the Taylor expansion of $\epsilon(q)$ and $K(q)$ near the equilibrium point $q_0$:
\bsubeqs\label{eq:expansions}
\beqa
K(q) &=& K(q_0) + K^\prime(q_0)\,(q-q_0)
+\text{O}\big( (q-q_0)^2 \big)\,,
\label{eq:expansions-K}\\[2mm]
\epsilon(q)-\mu_0\, q  &=& \epsilon^{\prime\prime}(q_0)\,(q-q_0)^2/2
+\epsilon^{\prime\prime\prime}(q_0)\,(q-q_0)^3/6
+\text{O}\big((q-q_0)^4 \big)\,.
\label{eq:expansions-epsilon}
\eeqa
\esubeqs
All coefficients in these expansions have Planck-scale values, for example,
$K(q_0) = 1/(16\pi G_{N})= (1/2)\,\EPlanck^2$
in terms of Newton's constant $G_{N}$ and the energy scale
$\EPlanck$ defined under \eqref{eq:VacuumEnergyOscillating-dimensionfull}.

We now consider the spatially flat FRW universe~\cite{Weinberg2008}
described by the Hubble expansion parameter $H\equiv (d a/d t)/a$
for scale factor $a(t)$ and
use the dimensionless variables $y\propto (q-q_{0})$ and $h\propto H$,
which have been rescaled with the Planck-scale parameters of the theory.
These two variables $y(\tau)$ and
$h(\tau)$ are governed by the following two coupled ordinary differential
equations (ODEs):
\bsubeqs\label{eq:Matter excluded1and2}
\beqa
\ddot{y} -\dot{y}\, h +2\, (1+y)\, \dot{h} &=&
-3 \big(1+w_{M}\big)\,\big[\dot{y}\, h
+ (1+y)\, h^2 - r_{V}  \big]\,.
\label{eq:Matter excluded1}
\\[2mm]
\dot{h} +2\, h^2   &=& r_{V}^\prime\,,
\label{eq:Matter excluded2}
\eeqa
\esubeqs
with the prime standing for differentiation with respect to $y$ and
the overdot for differentiation with respect to dimensionless
cosmic time $\tau$ (cosmic time $t$ in the corresponding Planckian units).
In the derivation of the above ODEs, the function $K(q)$ has been
assumed~\cite{KlinkhamerVolovik2008b} to be linear in $q$ for simplicity
[in terms of the coefficients of \eqref{eq:expansions-K}, one has
$q_0\,K^\prime(q_0) = K(q_0)$ and $K^{(n)}(q_0) = 0$ for $n \geq 2$].

The dimensionless vacuum energy density $r_{V}$ (vacuum energy
density $\rho_{V}$ in Planckian units) is taken to be given by
\begin{equation}
r_{V}(y)= \frac{1}{2}\,y^2 +\frac{2}{3}\,y^3 +\frac{1}{6}\,y^4\,,
\label{eq:DimensionlessVacEn2}
\end{equation}
which vanishes in the equilibrium state $y=0$, having chosen
$\mu=\mu_0$ in \eqref{eq:genMaxwellSolution}--\eqref{eq:genEinsteinEquation-rhoV}.
Later on, only the quadratic part of $r_{V}(y)$ will be relevant.
Equations \eqref{eq:Matter excluded1and2} and \eqref{eq:DimensionlessVacEn2}
lead to the rapid relaxation
\eqref{eq:VacuumEnergyOscillating-dimensionfull}, if the Universe starts
out with a nonequilibrium value of the charge,
$q_\text{initial}\neq q_0$ or $y_\text{initial} \ne 0$.
These Eqs.~\eqref{eq:Matter excluded1and2} and \eqref{eq:DimensionlessVacEn2}
are, in fact, identical to Eqs.~(5.2) and (5.3) in Ref.~\cite{KlinkhamerVolovik2008b},
to which the reader is referred for all details.

For the present analysis, it turns out to be useful to define
the following matter EOS parameter:
\begin{equation}
\kappa_{M} \equiv 4 - 3\,(1+w_{M})\,,
\label{eq:EOS}
\end{equation}
where $\kappa_{M}=0$ corresponds to matter with $T_{\alpha}^{\;\;\alpha}=0$,
for example, electromagnetic radiation (photons) or
ultrarelativistic massive particles (e.g., electrons and positrons).
Then, \eqref{eq:Matter excluded1} and \eqref{eq:Matter excluded2}
can be written as
\bsubeqs\label{eq:Matter excluded1and2new}
\beqa
\ddot{y} +3\, \dot{y}\, h  +2\, (1+y)\, r_{V}^\prime
&=&4\, r_{V} +\kappa_{M}\,\big[\dot{y}\, h
 + (1+y)\, h^2 - r_{V}  \big]\,,
\label{eq:Matter excluded1new}
\\[2mm]
\dot{h} +2\, h^2  - r_{V}^\prime &=& 0\,.
\label{eq:Matter excluded2new}
\eeqa
\esubeqs
The crucial observation, now, is that, for $\kappa_{M}(\tau)=0$,
there is a solution of the ODEs \eqref{eq:Matter excluded1new}
and \eqref{eq:Matter excluded2new},
where the vacuum energy density is exactly zero. This solution
corresponds to an FRW universe with ultrarelativistic matter present
but dark energy and cold dark matter (CDM) absent:
\bsubeqs\label{eq:PureRadiation}
\beqa
y(\tau)&=&0 \,,
\\[1mm]
h(\tau)&=&1/(2\, \tau)  \,,
\eeqa
\esubeqs
which, as said, holds for $\kappa_{M}(\tau)=0$.

Next, consider what happens when the model universe described
by \eqref{eq:PureRadiation} enters a phase at $t\sim t_\text{kick}$
for which $\kappa_{M}(t)\neq 0$.
Then, the vacuum variable $y$ becomes nonzero and a nonzero value of
the vacuum energy density emerges continuously.
Specifically, we consider a time $t_\text{kick}\gg \tPlanck$,
so that the corresponding dimensionless time
is large, $\tau_\text{kick}\gg 1$. At large $\tau$, the variable  $y(\tau)$ is always small
and one can make an expansion in terms of powers of $y$.
To first order in $y$ and $h^2$, one obtains the following ODEs
from \eqref{eq:Matter excluded1new} and \eqref{eq:Matter excluded2new}:
\bsubeqs\label{eq:Linearized1_y1andy2}
\beqa
\ddot y + 3\, h\, \dot{y} +  \omega^2\, y&=& \kappa_{M}\, h^2\,,
\label{eq:Linearized1_y1}
\\[2mm]
\dot{h} +2\, h^2 - y   &=& 0 \,,
\label{eq:Linearized1_y2}
\eeqa
\esubeqs
with  an implicit $\tau$ dependence for all three functions $y$, $h$,
and $\kappa_{M}$. Here, $\omega$ is the natural frequency of the
microscopic oscillations~\cite{KlinkhamerVolovik2008b},
which is given by $\omega=\sqrt{2}$  in Planckian units.

\section{Electroweak kick}
\label{sec:Electroweak-kick}

There are different regimes for the behavior of the
vacuum energy density obtained from \eqref{eq:Linearized1_y1andy2},
depending on the sharpness of the profile of the transition,
i.e., the width $\Delta\tau_\kappa$ of the function  $\kappa_{M}(\tau)$.
For the case of a smooth transition (that is, smooth on microscopic time scales,
$\Delta\tau_\kappa \gg 1/\omega \sim 1$), one may neglect the time derivatives
of $y$ in \eqref{eq:Linearized1_y1} to obtain:
\bsubeqs\label{eq:Linearized2_y1andy2}
\beqa
y&=&\kappa_{M}\, h^2/2\,,
\label{eq:Linearized2_y1}
\\[2mm]
y   &=& \dot{h} +2\, h^2\,,
\label{eq:Linearized2_y2}
\eeqa
\esubeqs
where the specific value $\omega^2=2$ has been reinstated in the first equation.
Eliminating $y$ from the above equations gives immediately the following solution
for $h(\tau)$:
\begin{equation}
h(\tau)=
\left[\, 2\int^\tau_0  d\tau'\,\Big(1-\kappa_{M}(\tau')/4 \Big)\right]^{-1}\,,
\label{eq:GeneralSolution}
\end{equation}
which holds for an arbitrary (smooth) function $\kappa_{M}(\tau)$
and has boundary condition $1/h(0)=0$, appropriate for
the standard hot big bang universe. Taking the square of
\eqref{eq:GeneralSolution},
the solution for $y(\tau)$  follows from \eqref{eq:Linearized2_y1}.

Now apply this result to the cosmological epoch of the electroweak
crossover~\cite{Csikor-etal1999}.
During the crossover, the Standard Model particles acquire masses and, as a
result, $w_{M}(t)$ deviates from 1/3. In principle, this deviation may be
enhanced by ``new physics'' at the TeV energy scale, which might be responsible
for the observed cold-dark-matter component of the present Universe by providing
a TeV--scale WIMP (weakly interacting massive particle).
According to the electroweak scenario of Ref.~\cite{ArkaniHamed-etal2000},
this new physics may have many particles ($n=1, \ldots ,N$)
with masses $M_n\sim \Eew \sim
1\;\text{TeV}$, which are created before and during the electroweak epoch.
Perhaps we will know from future particle-collider experiments
(for example, at the Large Hadron Collider of CERN) whether or not there
exists a TeV-scale WIMP responsible for the observed CDM.

Anyway, massive Standard Model particles
(and possible additional massive particles of new TeV--scale physics)
annihilate during the electroweak-crossover period and, afterwards,
the EOS parameter returns to its standard radiation-dominated value
$w_{M}=1/3$ [or $\kappa_{M}=0$], with the result that
the vacuum energy density is no longer perturbed.
In the epoch after the electroweak period when all perturbations have ceased,
the Hubble parameter \eqref{eq:GeneralSolution} is given by
\bsubeqs\label{eq:HubbleLateTime}
\beqa
h(\tau)&\approx&\frac{1}{2\,(\tau-\tau_0)}\,,
\\[1mm]
\tau_0&\equiv&\frac{1}{4}\,\int_0^\infty d\tau'~ \kappa_{M}(\tau')\,,
\eeqa
\esubeqs
for $\tau \gg \tau_0 \sim \tau_\text{ew}
\sim \EPlanck^2/\Eew^2 \sim 10^{30}$.

 From \eqref{eq:DimensionlessVacEn2} and \eqref{eq:Linearized2_y1},
the dimensionless and dimensionful vacuum energy densities
during the electroweak crossover  behave as follows:
\bsubeqs\label{eq:VacuumEnergySmooth}
\beqa
r_{V}(\tau)&=&(1/8)\,\kappa_{M}^2(\tau)\, h^4(\tau)\,,
\label{eq:VacuumEnergySmooth-rV}
\\[1mm]
\rho_{V}(t)&\propto& \kappa_{M}^2(t)\, H^4(t)\,,
\label{eq:VacuumEnergySmooth-rhoV}
\eeqa
\esubeqs
where only the quadratic part of \eqref{eq:DimensionlessVacEn2}
has been kept as $|y|\ll 1$ and where the precise numerical constant in
\eqref{eq:VacuumEnergySmooth-rhoV} depends on the microphysics
but can be expected to be of order unity~\cite{KlinkhamerVolovik2008b}.

Even though result \eqref{eq:VacuumEnergySmooth-rhoV} is similar to the
vacuum energy density estimate~\cite{Thomas-etal2009} from the
conformal anomaly,
$\rho_{V}(t)\sim \left<T_{\alpha}^{\;\;\alpha}\right>
\sim H^4(t)$, the mechanism of the emerging vacuum energy density
in \eqref{eq:VacuumEnergySmooth} is different.
The underlying theory~\cite{KlinkhamerVolovik2008b} of
result \eqref{eq:VacuumEnergySmooth} has, in fact, a gravitational
coupling parameter $K$ that depends on the vacuum variable, $K=K(q)$,
with Newton's constant recovered in the  $q=q_0$ equilibrium state,
$G_{N}=1/\big(16\pi\,K(q_{0})\big)$. Precisely this variability
$K(q)$ allows for a time-dependent vacuum energy density,
$\dot{\rho}_{V} \propto \dot{K}\,(\dot{H}+2H^2)$,
provided the expansion differs from that of a
radiation-dominated FRW universe with $H(t)=1/(2\,t)$ and $\dot{H}+2H^2=0$.

 From \eqref{eq:VacuumEnergySmooth-rhoV},
the magnitude of the vacuum energy density
at the crossover time \eqref{eq:t_ew} is given by
\begin{equation}
 \rho_\text{V,0}(\tew)
 \sim  H^4(\tew)
 \sim \tew^{-4}
 \sim \Eew^8/\EPlanck^4\,,
\label{eq:VacuumEnergySmoothEWtime}
\end{equation}
where $\kappa_{M}(\tew)$ has been assumed to be of order unity
and where, for later use, a suffix $0$ has been appended to
distinguish the ``classical'' result. This completes the first step
toward establishing a nonzero cosmological constant of the present Universe.
The second step is to make sure that the vacuum energy density generated
at $t\sim \tew \sim 10^{-12}\,\text{s}$ is not lost during the
remaining $10^{10}$ years.

\section{Subsequent evolution}
\label{sec:Subsequent_evolution}

The typical value of the vacuum energy
density \eqref{eq:VacuumEnergySmoothEWtime} emerging from
the electroweak crossover is comparable to the presently observed
value~\cite{Weinberg2008,Komatsu2008} of the vacuum energy
density.\footnote{An excellent description of the currently available data is,
in fact, given by the flat--$\Lambda$CDM model
(cf. Refs.~\cite{Weinberg2008,Komatsu2008}),
with an inhomogeneous cold-dark-matter component (EOS parameter $w_\text{CDM}=0$)
and a perfectly homogeneous and time-independent vacuum energy density component
($w_{V}=-1$), which corresponds to Einstein's cosmological
constant $\Lambda$.\label{ftn:LambdaCDM}}
As mentioned in Sec.~\ref{sec:Introduction}, this suggests a possible explanation
of the triple cosmic coincidence according to the electroweak scenario
discussed in Ref.~\cite{ArkaniHamed-etal2000}. But,
for this explanation to work, we need a mechanism to stabilize
the vacuum energy density after the electroweak crossover.

At the moment, we do not have a complete theory which describes the
\emph{irreversible} dynamics of the quantum vacuum. The classical equations
of $q$--theory~\cite{KlinkhamerVolovik2008a} describe only the
\emph{reversible} classical dynamics of the vacuum. One needs to extend
$q$--theory to the quantum domain, in order to incorporate the dissipative
relaxation of the vacuum energy density due to the quantum effect of matter field
radiation (matter quanta emission).

Awaiting the definite theory of the quantum vacuum,
the following model equation can be used for a rough estimate:
\begin{equation}
\dot{\rho}_{V} = - \Gamma(t)\,\big[\rho_{V}(t)- \rho_\text{V,0}(t) \big]\,.
\label{eq:relaxation_equation}
\end{equation}
Here, $\rho_\text{V,0}(t)$  is the ``bare'' vacuum energy density driven
by the kick, which, according to result \eqref{eq:VacuumEnergySmooth-rhoV} of
the classical $q$--theory, is given by
\begin{equation}
 \rho_\text{V,0}(t) \propto   \kappa_{M}^2(t)\, H^4(t)\,,
\label{eq:classical_value}
\end{equation}
and $\Gamma(t)\geq 0$ in \eqref{eq:relaxation_equation} is the rate
at which the ``surplus'' vacuum energy density is dissipated into particles.

Particle production occurs when the background spacetime is changing on a
timescale comparable to the particle Compton time~\cite{BirrellDavies1980},
which implies different particle production rates for different
cosmological epochs. In the epoch before the electroweak crossover,
matter consists of ultrarelativistic particles (radiation)
with EOS parameter $\kappa_{M}=0$ and, thus,
there is no ``external force'' to drive the vacuum energy density.
Rapid oscillations with frequency $\omega \sim \EPlanck$
lead to the decay of the vacuum energy density with the rate
$\Gamma\sim \omega \sim \EPlanck$~\cite{Starobinsky1980,Vilenkin1985}.
As a result, \eqref{eq:relaxation_equation} gives exponential decay
\eqref{eq:VacuumEnergyQuantum-dimensionful} of the
vacuum energy density to a zero value.
The model universe rapidly approaches the stage with pure
radiation, evolving as in \eqref{eq:PureRadiation}.

During the electroweak crossover, the EOS parameter $ \kappa_{M}(t)$
in \eqref{eq:classical_value} deviates from zero, which drives the
vacuum energy density \eqref{eq:relaxation_equation} away from
zero towards a positive value. The change of the vacuum energy density
during the crossover results in the emission of particles.
The radiation rate  $\Gamma(t)$ is concentrated in the crossover period,
because after the crossover the model universe returns to radiation-dominated
expansion without particle production. The decay rate $\Gamma(t)$ is,
therefore, peaked at $t\sim \tew$,
\beq
\Gamma(t)\;\big|_{t \ll \tew \;\vee\; t \gg \tew}  \ll \Gamma(\tew) \sim 1/\tew\,,
\label{eq:Gamma-peak-value}
\eeq
where the maximal value $1/\tew$ will be derived shortly. Note that
the maximal rate $\Gamma(\tew) \sim \Eew^2/(\hbar\,\EPlanck)$
goes to infinity for $\hbar\to 0$
and fixed energy $\Eew^2/\EPlanck$,
so that \eqref{eq:relaxation_equation} reproduces the classical result,
$\rho_{V}(t)\to \rho_\text{V,0}(t)$. In fact, this particular classical
limit corresponds to the hydrodynamic limit in fluid dynamics; cf. the
section on ``second viscosity'' in Ref.~\cite{LandauLifshitz-Fluid-mechanics}.
Further remarks on the heuristics of the vacuum dynamics equation
\eqref{eq:relaxation_equation} will be presented in the
paragraph starting a few lines after \eqref{eq:FinalrhoV}.

The estimate for the maximal value of the decay rate
in \eqref{eq:Gamma-peak-value} can be obtained as follows.
Start from the observation~\cite{ZeldovichStarobinsky1977} that,
for an FRW universe with appropriate boundary conditions~\cite{DobadoMaroto1999},
the number of particles created per unit of time and per unit of volume
is given by $\dot{n} \propto R^2$, where $R$ is the Ricci curvature scalar.
For an FRW universe with pure radiation, the Ricci scalar
$R\propto (\dot{H}+2H^2)$ vanishes and there is no particle production.
As mentioned before,
this is the reason why the radiation rate $\Gamma(t)$ is peaked
in the crossover period.\footnote{Matter radiation must also vanish in a
de-Sitter spacetime, where no relaxation of the vacuum energy density
is expected. For a discussion of the controversies
concerning the stability of de-Sitter spacetime, see, e.g.,
Refs.~\cite{Starobinskii1979,StarobinskyYokoyama1994,GarrigaTanaka2007,
TsamisWoodard2007,Busch2008,Volovik2008}.\label{ftn:deSitter}}
In the period of the electroweak crossover, one
has $R^2( \tew) \sim \dot{H}^2( \tew)  \sim H^4(
\tew)\sim \rho_{V}( \tew)$.
Particles created~\cite{BirrellDavies1980} during this period
have a Compton time of order $\tew$ and, thus, a
characteristic energy of order $E\propto 1/ \tew$.
The only known elementary particles whose energy $E$ can be of order
$1/ \tew \sim  \Eew^2/\EPlanck \sim \text{meV}$
are massless gravitons and massive neutrinos, some of
whose masses~\cite{Amsler-etal2008} may be comparable
with $1/ \tew$ (all the other particles of the Standard
Model have larger masses, including the photon which gets an
effective mass in the cosmic plasma).
During the electroweak-crossover period, the radiated energy
per unit of time and per unit of volume
is then $\dot{\rho}_{V}  \propto - E\,\dot{n} \propto -\rho_{V}/
\tew$, giving $\Gamma(\tew) \sim 1/\tew$
for the decay rate entering \eqref{eq:relaxation_equation}
and delivering the announced estimate \eqref{eq:Gamma-peak-value}.

Now, the solution of \eqref{eq:relaxation_equation} is given by
\beq
\rho_{V}(t)=  \int_0^{t} dt' \;\Gamma(t')\, \rho_\text{V,0}(t')\;
                   \exp\left[ -\int_{t'}^{t} dt''\;\Gamma(t'')\right]\,,
\label{eq:general-solution-rhoV}
\eeq
for boundary condition $\rho_{V}(0)=0$,
which is reasonable for times $t$ well after the Planck era.
Since $\Gamma(t)$ is concentrated in the crossover period and has
peak value \eqref{eq:Gamma-peak-value}, the solution
\eqref{eq:general-solution-rhoV} gives
$\lim_{t \rightarrow \infty}\,\rho_{V}(t)\sim \rho_\text{V,0}(\tew)$.
For very late times, $t\gg \tew$,
one thus obtains that the vacuum energy density approaches
the following positive and time-independent value:
\beqa
  \rho_{V}(t)\;\big|_{t \gg  \tew}
  &\sim& \rho_{V}(\tew)
  \sim \rho_\text{V,0}(\tew)
\sim \Eew^8/\EPlanck^4\,,
\label{eq:FinalrhoV}
\eeqa
where  \eqref{eq:VacuumEnergySmoothEWtime} has been  used in the last step.
The final result \eqref{eq:FinalrhoV} is comparable to the measured value
of the cosmological constant, as shown in \eqref{eq:vac_energy_at_present}.

The heuristics of the obtained nonzero remnant vacuum energy density
is as follows.
The quantity $\Gamma(t)$ in \eqref{eq:relaxation_equation} can be
interpreted as the inverse of the instantaneous response time $\theta(t)$
of the vacuum energy density $\rho_{V}(t)$ to an ``external perturbation.''
Here, the external perturbation \eqref {eq:classical_value} comes from the
``kick'' in $\kappa_{M}(t)$, which is assumed to happen at $t\sim \tew$
and to have a full width at half maximum $\Delta t_\kappa\sim \tew$.
Moreover,  $\Gamma(t)\equiv 1/\theta(t)$ is taken to have a width
$\Delta t_\Gamma$, which is comparable to or larger than the duration of the kick,
$\Delta t_\Gamma \gtrsim \Delta t_\kappa$.
\mbox{\emph{A priori},} there are then two possibilities.
First, the typical response time $\theta$ is short ($\theta \ll \Delta t_\kappa$),
which implies that the vacuum energy density $\rho_{V}(t)$ can follow
the kick in $\kappa_{M}(t)$ and that
$\rho_{V}(t)$ can recover a near-zero value,
as $\kappa_{M}(t)$ drops to zero for $t\gg \tew$.
Second, the typical response time $\theta$ is relatively long
($\theta\gtrsim \Delta t_\kappa$), which implies that the vacuum energy density
$\rho_{V}(t)$ cannot keep up with $\kappa_{M}(t)$, as the latter
drops to zero, and that a nonzero asymptotic value of $\rho_{V}$
remains. According to \eqref{eq:Gamma-peak-value}, this second type
of behavior occurs for the case considered,
with $\theta \sim \Delta t_\kappa \sim \tew$,
and a nonvanishing asymptotic value of $\rho_{V}(t)$
follows from the general solution \eqref{eq:general-solution-rhoV}.
In short, the nonzero remnant vacuum energy density
\eqref{eq:FinalrhoV} is a \emph{time-lag effect}, because the response
(relaxation) time of the vacuum energy density is of the same order of
magnitude as the duration of the kick.\footnote{In principle,
the same time-lag (freezing) mechanism may work for the scenario of
Ref.~\cite{Thomas-etal2009}, where a vacuum energy density $\rho_{V}\propto
H^4(t)$ emerges due to the conformal anomaly. During the electroweak crossover,
the number of massless fields contributing to the anomaly changes, which
results in a kick of the vacuum energy density.  In turn, this gives rise to
matter radiation, which leads to the stabilization of the
vacuum energy density at a value of the order of \eqref{eq:FinalrhoV}.}

After the electroweak crossover, further perturbations
of the vacuum energy density occur during the QCD confinement transition at
a typical temperature $T\sim E_\text{QCD} \sim 10^2\;\text{MeV}$
and the epoch following the moment of radiation-matter equality,
when the radiation-dominated effective EOS parameter $w_{M}=1/3$
changes to the matter-dominated parameter $w_{M}=0$.
[The moment of radiation-matter energy density equality happens to be close
to the epoch of recombination with $T \sim T_\text{rec} \sim 10^{-1}\;\text{eV}$
and this energy scale will be used for definiteness.] The first-mentioned
perturbation of the vacuum energy density by the QCD confinement transition
(see, e.g., Fig.~19.3 in Ref.~~\cite{Amsler-etal2008} for the change in the
number of relativistic degrees of freedom)  can be expected to give a
change of the order of $H^4(t_\text{QCD})\sim (E_\text{QCD}^2/\EPlanck)^4$,
which is negligible compared to the present value of $\Lambda$
according to \eqref{eq:vac_energy_at_present}.
The second perturbation of the vacuum energy density acts during the whole
matter-dominated era. However, the resulting change of the vacuum energy density
can be expected not to exceed a value of order
$H^4(t_\text{rec})\sim (T_\text{rec}^2/\EPlanck)^4$, which is, again, many
orders of magnitude smaller than the present value of $\Lambda$ and can
be neglected.

Turning the argument of the preceding paragraph around, it would seem that
the suggested electroweak explanation \eqref{eq:vac_energy_at_present}
of the present value of $\Lambda$ would rule out
(leave no room for) similar crossover effects at much higher temperature
$T_\star \gg \Eew \sim $ \text{TeV},
the expected remnant vacuum energy density $H^4(t_\star)$
being much larger than $H^4(\tew)$.
This conclusion, if correct, may be consistent with the
picture~\cite{ArkaniHamed-etal2000} of having only two fundamental energy scales,
$\Eew$ and $\EPlanck$, without unification of the
Standard Model gauge group at an intermediate grand-unification
energy scale~\cite{KlinkhamerVolovik2005,Kawamura2009}.

\section{Discussion}\label{sec:Discussion}

The $q$--theory approach~\cite{KlinkhamerVolovik2008a}
to gravitational effects of the quantum vacuum suggests at least two types
of behavior for the evolution of the vacuum energy density,
each based on solutions of the $q$--theory dynamical equations
and their modifications due to dissipative effects from matter radiation.
For the first type of
solution~\cite{KlinkhamerVolovik2008b,KlinkhamerVolovik2008c},
the model universe is vacuum dominated with,
according to  \eqref{eq:VacuumEnergyOscillating-dimensionfull},
the vacuum energy density $\rho_{V}(t)$ relaxing as $1/t^2$
from its  natural Planck-scale value at early times when the system is
far from equilibrium to a naturally small value at late times when the system is
close to equilibrium. [Quantum effects (e.g., the emission of matter quanta
caused by the rapid oscillations of the vacuum state)
make the relaxation even faster,
as shown by \eqref{eq:VacuumEnergyQuantum-dimensionful}.]
This essentially solves the main cosmological problem
(but with the \emph{caveat} mentioned in Sec.~\ref{sec:Dynamical_equations}):
the present vacuum energy density is small compared to Planck-scale values
simply because the age of our Universe happens to be large
compared to Planck-scale values.
However, it leaves the following question: why does not the vacuum energy density
relax completely to zero as $t\rightarrow \infty$?

In order to answer this last question,
we presented a second type of solution in which the vacuum energy
density has already relaxed to  zero after the initial disturbance
in the very early universe and a nonzero value reemerges only  after a ``kick''
generated by nonrelativistic matter during the epoch of the electroweak crossover.
(These nonrelativistic particles consist of Standard Model particles and possibly
thermal relics from new physics at the $\text{TeV}$ scale, as
discussed in Sec.~\ref{sec:Electroweak-kick}.) In the process, a
nonoscillating vacuum energy density is generated, which starts to decay
after the kick. Such a behavior  emerges during the electroweak period, because
in this epoch the matter EOS parameter $w_{M}(t)$ deviates from the
radiative value $w_{M}=1/3$. Quantum effects now lead to a
stabilization of the vacuum energy density at the level indicated
by \eqref{eq:FinalrhoV}, which reproduces the expression suggested
previously by Arkani-Hamed \emph{et al.}~\cite{ArkaniHamed-etal2000}.

It was assumed in the reasoning leading up
to \eqref{eq:FinalrhoV} that there was no real phase
transition at cosmic time $\tew$. Instead, there was taken to be a
crossover at a temperature $T_\text{ew} = \text{O}(10^2\;\text{GeV})$,
which does not give a change of order $T_\text{ew}^4$
in the vacuum energy density as a genuine phase transition would do.
The absence of a real electroweak phase transition is by now well
established~\cite{Csikor-etal1999}, at least, in the framework of
the Standard Model of elementary particle physics
(the numerical value of the crossover temperature is
estimated~\cite{Csikor-etal1999}
as $T_\text{ew} \sim 300\;\text{GeV}$ for $m_\text{Higgs}\sim 150\;\text{GeV}$).
The new physics at the $\text{TeV}$ scale mentioned in the previous
paragraph and Sec.~\ref{sec:Electroweak-kick}
is assumed not to affect the nature of the electroweak crossover.
But the massive relic particles of the new physics can make a significant
contribution to the EOS parameter $\kappa_{M}(t)$ and can also
increase the numerical value of the effective energy scale $\Eew$,
thereby augmenting the magnitude of the estimated dark energy
\eqref{eq:classical_value}--\eqref{eq:general-solution-rhoV}
and bringing the theoretical value \eqref{eq:vac_energy_at_present}
closer to the observed value~\cite{Weinberg2008,Komatsu2008} of approximately
$(2\;\text{meV})^4$.

The electroweak scenario of Ref.~\cite{ArkaniHamed-etal2000} may
solve part of the triple cosmic coincidence
puzzle, as the same order of magnitude follows naturally for
the cold-dark-matter density and the radiation density in the present epoch.
Combined with the argument for the effective cosmological
constant \eqref{eq:vac_energy_at_present} of the present article,
this suggests that $\text{TeV}$--scale physics may be responsible for the
triple coincidence of vacuum, matter, and radiation energy densities
in the present Universe (perhaps even a quintuple coincidence if also
the baryon and neutrino energy densities are considered~\cite{ArkaniHamed-etal2000}).

For the present epoch, the vacuum energy density would be essentially
time-independent according to \eqref{eq:general-solution-rhoV}
and, observationally, the corresponding universe would be indistinguishable
from the one of the $\Lambda$CDM model (cf. Footnote~\ref{ftn:LambdaCDM}).
But, theoretically, we would have gained in understanding the
magnitude of the cosmological ``constant'' $\Lambda$
as given by \eqref{eq:vac_energy_at_present}, in addition
to explaining the triple or quintuple cosmic coincidence mentioned above.

\section*{\hspace*{-4.5mm}ACKNOWLEDGMENTS}
\noindent
It is a pleasure to thank A.A. Starobinsky and A.R. Zhitnitsky
for useful discussions on basic physics issues
and the referee for helpful comments on an earlier version of this article.
GEV is supported in part by the Academy of Finland,
Centers of Excellence Program 2006--2011,
the Russian Foundation for Basic Research (Grant No. 06--02--16002--a),
and the Khalatnikov--Starobinsky leading scientific school (Grant No. 4899.2008.2).

\end{document}